\documentclass[preprint, showpacs, preprintnumbers, amsmath,amssymb]{revtex4}

\usepackage{graphicx}
\usepackage{dcolumn}
\usepackage{tensor}
\usepackage{fontenc}

\begin{document}

\title{ Thermodynamical properties of interacting holographic dark energy model with apparent horizon }
\author{Bin Liu}
\author{Xian-Ru Hu}
\author{Jian-Bo Deng}
\email{dengjb@lzu.edu.cn} \affiliation{
Institute of Theoretical Physics, Lanzhou University \\
Lanzhou 730000, People's Republic of China}

\date{\today}

\begin{abstract}

We have investigated the thermodynamical properties of the universe with dark energy.
It is demonstrated that in a universe with spacial curvature the natural choice for 
IR cutoff could be the apparent horizon radius. We shown that any interaction of pressureless 
dark matter with holographic dark energy, whose infrared cutoff is set by the apparent horizon radius, 
implying a constant effective equation of state of dark component in a universe. 
In addition we found that for the static observer in space, 
the comoving distance has a faster expansion than the apparent horizon radius with any spatial curvature. 
We also verify that in some conditions the modified first law of thermodynamics 
could return to the classic form at apparent horizon for a universe filled with dark energy and dark matter. 
Besides, the generalized second law of thermodynamics is discussed in a region enclosed by the apparent horizon.

\pacs{98.80.-k, 95.36.+x}
\textbf{Keywords}:{ holographic dark energy, interaction.}

\end{abstract}
\maketitle

\section{introduction}

From the cosmological observations \cite{f1}\cite{f2}\cite{f3}\cite{f4}\cite{f5}, 
it is generally accepted in the astrophysics community that the universe is expanding in 
an accelerated manner due to some kind of negative-pressure form of matter known as 
dark energy \cite{f6}. Therefore, in order to explain this bizarre phenomenon, 
various models of dark energy have been proposed. 
The most naive dark energy candidate is the cosmological constant \cite{f7}, 
conventionally associated with the energy of the vacuum with constant energy density and pressure, 
and an equation of state $\omega=-1$. However, it is confronted with two fundamental problems: 
the fine-tuning problem and the cosmic coincidence problem \cite{f8}. 
To alleviate the drawbacks, cosmologists and particle physicists have proposed dynamical dark energy models, 
including minimally coupled scalar fields, vector fields or even modifications to General Relativity on cosmological scales \cite{f9}\cite{f10}, 
such as those associated with extra-dimensions \cite{f11}\cite{f12}\cite{f13} and f(R) theories \cite{f14}\cite{f15}\cite{f16}\cite{f17}.\\
\indent Anther interesting attempt for probing the nature of dark energy within the framework of 
quantum gravity is the so-called "HDE" (Holographic Dark Energy), proposal motivated from the holographic hypothesis \cite{f18}. 
In ref.\cite{f19} they suggested a relationship between the ultraviolet (UV) and the infrared (IR) cutoffs 
due to the limit set by the formation of a black hole, $L^3\Lambda^4 \leq L M^2_{pl}$, 
where $\Lambda$ as the UV cutoff is closely connected to the quantum zero-point energy density, 
$L$ is the size of the region and $M_{pl}=\frac{1}{\sqrt{8 \pi G}}$ is the reduced Planck mass. 
Applying this idea to the whole universe, the largest IR cutoff $L$ is chosen by saturating the inequality, 
one can get the holographic dark energy density\\
\begin{equation}\label{eq1}
\rho_d=3c^2M^2_{pl}L^{-2}
\end{equation}
where $3c^2$ is a numerical constant characterizing all of the uncertainties of the theory, 
and its value can only be determined by observations. 
If we take $L$ as the Hubble horizon $H^{-1}$, 
the resulting $\rho_d$ is comparable to the observational density of dark energy \cite{f20}\cite{f21}. 
However, Hsu \cite{f22} pointed out that this does not lead to an accelerated universe. 
Subsequently, in \cite{f23} Li suggested that the IR cut-off $L$ should be taken as the size of the future 
event horizon of the universe. For a comprehensive list of references concerning HDE, we refer to \cite{f24}\cite{f25}
\cite{f26}\cite{f27}\cite{f28}\cite{f29} e.g.\\
\indent Following above discussion, although fundamental for our understanding of the universe, 
the nature of such previously unforeseen energy remains completely open questions, 
especially in the theoretical aspect. 
One of the important questions concerns the thermodynamical behavior of the accelerated expanding universe driven by dark energy. 
In history, the connection between thermodynamics and gravity has been focused for several times \cite{f30}\cite{f31}. 
This connection implies that the thermodynamics might shed some light on the properties of dark energy, 
so it gives strong motivation to study thermodynamics in the accelerating universe. \\
\indent In this work we would like to study the dark energy from thermodynamical perspective. 
In \cite{f32} the interaction between dark energy and dark matter is taken into account, 
they proposed a constant ratio of the energy densities of both components by the Friedmann equation and solve the coincidence problem. 
Based on this, according to the interaction parameter, 
taking the apparent horizon radius as the infrared cutoff, 
we demonstrate that the effective equation of state of dark component could also be proportional to 
the constant in a universe with any spatial curvature. Besides, motivated by the idea in \cite{f33}, 
we show that for the static observer in space, the comoving distance has a faster expansion than the apparent horizon radius. 
This implies that with the accelerating expansion of the universe, 
the contents enclosed by the apparent horizon disappear continuously and crossing the horizon, 
since the distribution of dark component is evenly across space. 
Therefore, it is of great interest to investigate whether the first law of thermodynamics is fulfilled on the apparent horizon, 
and the validity of the generalized second law of thermodynamics in the accelerating universe driven by dark energy.\\
\indent The paper is organized as follows. In section II, we study the interacting HDE with 
apparent horizon as an IR cutoff and propose the effective conservation equation. 
Then we discuss the first law of thermodynamics on the apparent horizon in an accelerating universe 
with commoving coordinate system in section III. In section IV, 
we examine the validity of the generalized second law of thermodynamics in a region enclosed by the apparent horizon. 
The summary and discussion are presented in the last section.

\section{INTERACTING HOLOGRAPHIC DARK ENERGY WITH APPARENT HORIZON}
\indent In this section we would describe the interacting holographic dark energy 
in a universe with each spatial curvature. First we consider a homogenous and isotropic 
universe which is described by the line element
\begin{equation}\label{eq2}
\mathrm{d}s^2=-\mathrm{d}t^2+a^2(t) \left[\frac{\mathrm{d}r^2}{1-kr^2}+r^2\left(\mathrm{d}\theta^2+sin^2\theta \mathrm{d}\phi^2\right)\right]
\end{equation}
Here $k$ denotes the curvature of space with $k=-1, 0, 1$ corresponding to open, flat, and closed universes, respectively. 
From ref.\cite{f34}, a simple calculation gives the dynamical apparent horizon radius for the FRW universe
\begin{equation}\label{eq3}
R_A=\frac{1}{\sqrt{H^2+k/a^2}}
\end{equation}
The apparent horizon may be considered as a causal horizon for a dynamical spacetime, 
so one can associate a gravitational entropy and surface gravity to it. 
The corresponding Friedmann equation reads
\begin{equation}\label{eq4}
(H^2+\frac{k}{a^2})=\frac{1}{R_A^2}=\frac{8\pi G}{3}(\rho_d+\rho_m)
\end{equation}
where $H=\dot{a} / a$ is the Hubble parameter, $\rho_d$ is the holographic dark energy density from eq.(\ref{eq1}), 
and $\rho_m$ is the energy density of matter. It is noted that the total energy density $\rho=\rho_d+\rho_m$ satisfies a conservation law
\begin{equation}\label{eq5}
\dot{\rho}+ 3H(\rho+p_d)=0
\end{equation}
However, since the interaction between dark energy and dark matter is taken into account, 
$\rho_d$ and $\rho_m$ do not satisfy the independent conservation laws. Instead they satisfy two continuous equations
\begin{equation}\label{eq6}
\dot{\rho}_d+ 3H \rho_d(1+\omega_d)=-Q
\end{equation}
\begin{equation}\label{eq7}
\dot{\rho}_m+ 3H \rho_m=Q
\end{equation}
here $\omega_d=p_d/\rho_d$ denotes the equation of state parameter (EoS) of HDE, 
and $Q$ is the interaction term which can be an arbitrary function of cosmological parameters. 
It is important to note that the continuity equations imply that the interaction term should be a 
function of a quantity with units of inverse of time multiplied with the energy density, 
and the ideal interaction term must be motivated from the theory of quantum gravity. 
Therefore, without loss of generality, we rely on pure dimensional basis for choosing an interaction $Q$ 
and write the interaction as $Q=\Gamma \rho_d$. By introducing a coupling constant $b^2$, 
we can present the interaction term as
\begin{equation}\label{eq8}
Q=\Gamma \rho _d=\left\{
\begin{array}{l}
3Hb^{2}\rho _{d} \\
3Hb^{2}u\rho _{d} \\
3Hb^{2}(1+u)\rho _{d}\\
\end{array}
\right.
\end{equation}
With $r=\rho_m /\rho_d$ stands for the ratio of energy densities. 
The effective equations of state for dark energy and matter are defined by \cite{f35}
\begin{equation}\label{eq9}
\omega_d^{eff}=\omega_d+\frac{\Gamma}{3H},\ \
\omega_m^{eff}=-\frac{1}{r}\frac{\Gamma}{3H}
\end{equation}
\indent In \cite{f32}, they show that any interaction of dark matter with holographic dark energy 
implies an accelerated expansion and $r$ is a constant. Thus that with eq.(\ref{eq8}), 
we find $\frac{\Gamma}{3H}\propto Const$. 
Making use of (\ref{eq9}) in (\ref{eq6}) and (\ref{eq7})
\begin{equation}\label{eq10}
\dot{\rho}_d+ 3H \rho_d(1+\omega_d^{eff})=0
\end{equation}
\begin{equation}\label{eq11}
\dot{\rho}_m+ 3H \rho_m(1+\omega_m^{eff})=0
\end{equation}
From eqs.(\ref{eq3}) and (\ref{eq5}), we can obtain
\begin{equation}\label{eq12}
-\frac{\dot{R}_A}{R_A^3}=H\left(\dot{H}-\frac{k}{a^2}\right)=\frac{4\pi G}{3}(\dot{\rho}_d+\dot{\rho}_m)=-4\pi GH(\rho_d+\rho_m+p_d)
\end{equation}
Differentiating eq.(\ref{eq1}) with respect to cosmic time and using (\ref{eq12}) yields
\begin{equation}\label{eq13}
\dot{\rho}_d=-2\rho_d\frac{\dot{R}_A}{R_A}=-3c^2H\rho_d(1+r+\omega_d)
\end{equation}
Substituting (\ref{eq13}) into (\ref{eq10}), we get
\begin{equation}\label{eq14}
\omega_d^{eff}=c^2(1+r+\omega_d)-1
\end{equation}
Following the same logic in \cite{f36}, we rewrite the eq.(\ref{eq6})
\begin{equation}\label{eq15}
3H(1+\omega_d)=-\frac{\dot {\rho}_d}{\rho_d}-\Gamma
\end{equation}
and eq.(\ref{eq7})
\begin{equation}\label{eq16}
\dot {r}=-r \frac{\dot {\rho}_d}{\rho_d}-3Hr+\Gamma
\end{equation}
Eliminating the $ \dot {\rho}_d/\rho_d$ in above two equations, we reach
\begin{equation}\label{eq17}
\dot {r}=(1+r)\Gamma+3Hr\omega_d
\end{equation}
For the case of the ratio of the energy densities is a constant, 
the equation of state parameter $\omega_d$ can be calculated
\begin{equation}\label{eq18}
\omega_d=-\left(1+\frac{1}{r}\right)\frac{\Gamma}{3H}
\end{equation}
Using (\ref{eq18}), (\ref{eq14}) gives 
\begin{equation}\label{eq19}
\omega_d^{eff}=c^2\left[1+r-\left(1+\frac{1}{r}\right)\frac{\Gamma}{3H}\right]-1
\end{equation}
The above expression represents the effective equation of state for the holographic dark energy interacting with dark matter. 
An interesting case arises when interaction is absence, then eq.(\ref{eq19}) gives 
$\omega_d=0$ and $\omega_d^{eff}=c^2(1+r)-1$. If $c=1$, $\omega_d^{eff}=r>0$. 
One could make a constraint to the constant $c$ to obtain a current accelerated expansion universe 
and to protect positivity of dark matter energy density. 
When an interaction is taken into account, the interaction parameter $\frac{\Gamma}{3H}$ 
together with constant $c$ could determine the region of effective equation of state parameter. 
An interaction is necessarily accompanied by an equation of state $\omega_d^{eff}<0$. 
Base on the above discussion, the existence of an interaction has another interesting consequence. 
When the infrared cutoff is set by the apparent horizon radius, it is found that $\omega_d^{eff}\propto Const$ 
and $\omega_m^{eff}\propto Const$ . Thus we could modified the continuous equations
\begin{equation}\label{eq20}
\rho_d \sim a^{-3(1+\omega_d^{eff})},\ \
\rho_m \sim a^{-3(1+\omega_m^{eff})}
\end{equation}
Here $a$ denotes the scale factor. It is shown that the definition of the coupling function 
$\Gamma$ together with the ratio of the energy densities fundamentally lead the relation 
between $\rho$ and $a$ to be a simple one. Next, we would verify that using eq.(\ref{eq20}) 
the first law of thermodynamics with comoving coordinate system on the apparent horizon could return to the classic form.

\section{FIRST LAW OF THERMODYNAMICS ON THE APPARENT HORIZON}

We first review the model of expanding universe. 
There are decisive evidences that our observable universe evolves adiabatically after inflation in a comoving volume. 
This means there is no energy-momentum flow between different patches of the observable universe then 
the universe keeps homogeneous and isotropic after inflation, which is why we can use FRW geometry to describe 
the evolution of the universe. In an adiabatically evolving universe, term $T\mathrm{d}S$ equals zero, 
the first law of thermodynamics in a comoving volume reads 
\begin{equation}\label{eq21}
\mathrm{d}U=-p\mathrm{d}V
\end{equation}
Here $U=\Omega_k \rho a^3$ and $V=\Omega_k a^3$ stand for the energy and the physical volume \cite{f37}, 
and $\Omega_k$ is a factor related to the spatial curvature
\begin{equation}\label{eq22}
\left\{
\begin{array}{lll}
k=-1 &  \rightarrow & \Omega_{-1}=\frac{e^{2\pi}-1/e^{2\pi}}{2}\pi-2\pi^2\\
k=0 &  \rightarrow & \Omega_{0}=\frac{4\pi}{3} \\
k=1 &  \rightarrow & \Omega_{1}=2\pi^2\\
\end{array}
\right.
\end{equation}
It is noted that eq.(\ref{eq21}) equals the continuity equation 
because the distribution of dark component is evenly across space. 
In the case of the interacting holographic dark energy, making use of the effective equation of state for dark energy and matter, 
the integral relation between $\rho$ and $a$ is given by eq.(\ref{eq20})\\
\indent In light of the discussion in \cite{f37}, 
we would concentrate our attention on the thermodynamical properties on the apparent horizon. 
It is a fact that for the static observer in space, the universe he/she observed is the space in a region enclosed 
by the apparent horizon. Compared with the model discussed above, 
it is of interest to see whether the expansion of the apparent horizon radius is consistent with the commoving distant. 
In other words, if the comoving distance has a faster or slower expansion, some energy should flow through the apparent horizon.\\
\indent Here we assume that at a certain beginning time, $t=t_0$, the comoving distance is same with the apparent horizon radius
\begin{equation}\label{eq23}
a^0 \xi=R_A^0
\end{equation}
where $a \xi$ is the comoving distant and $\xi\propto Const$ stand for the comoving coordinate. 
Making use of eqs.(\ref{eq3}) and (\ref{eq23}), 
a simple calculation shows that the change ratio of both of them is
\begin{equation}\label{eq24}
\frac{(a\xi)'}{(R_A)'}=\frac{H^2+\frac{k}{a^2}}{H^2+\frac{k}{a^2}-\frac{\ddot{a}}{a}}>1
\end{equation}
where a prime presents the derivative with respect to cosmic time. 
That means the comoving distance has a faster expansion than the apparent horizon radius with any spatial curvature. 
Because the distribution of total matter and energy matter is evenly across space, 
energy is conserved in the commoving volume according to the black hole thermodynamics. 
However, it is different that for the static observer in space there should be not only an energy density 
decrease but also some energy outflow the apparent horizon synchronously (FIG.1).\\
\indent In this paper, if we define $E_A^0=\rho V_1=\Omega_k(R_A^0)^3\rho^3$ 
as the total energy content of the universe inside a 3-sphere of radius $R_A^0$ at $t_0$, 
and $E_a^0=\rho V_2=\Omega_k(a^0 \xi)^3\rho^3$ is the energy enclosed by the comoving 
distance $a^0 \xi$ at $t_0$ respectively, then after a time $\Delta t$, $t_0 \rightarrow t_1$, 
the energy crossing the apparent horizon is
\begin{equation}\label{eq25}
\Delta E_A=E_A^1-E_a^1
\end{equation}
for the choice of spatially flat case
\begin{equation}\label{eq26}
\Delta E_A=\Omega_0\left[(R_A^1)^3-(a^1\xi)^3\right]\rho^1=\frac{4\pi}{3}\left[(R_A^1)^3-(a^1\xi)^3\right]\rho^1
\end{equation}
using $\rho^1=\rho^0+\Delta \rho$, we get
\begin{equation}\label{eq27}
\Delta E_A=\frac{4\pi}{3}\left[(R_A^1)^3-(a^1\xi)^3\right](\rho^0+\Delta \rho)=\frac{4\pi}{3}(R_A^1)^3\rho^0-\frac{4\pi}{3}(a^1\xi)^3\rho^1+\frac{4\pi}{3}(R_A^1)^3\Delta \rho
\end{equation} 
Since $\rho=\rho_d+\rho_m$, and $\rho$ is proportional to the $a^{-3(1+\omega^{eff})}$ in eq.(\ref{eq20})
\begin{eqnarray}\label{eq28}
\Delta E_A&=&\left[\frac{4\pi}{3}(R_A^1)^3\rho^0_m-\frac{4\pi}{3}(a^0\xi)^3\rho^0_m\frac{(a^0)^{1+\omega_m^{eff}}}{(a^1)^{1+\omega_m^{eff}}}+\frac{4\pi}{3}(R_A^1)^3\Delta \rho_m\right] \notag\\
&+&\left[\frac{4\pi}{3}(R_A^1)^3\rho^0_d-\frac{4\pi}{3}(a^0\xi)^3\rho^0_d\frac{(a^0)^{1+\omega_d^{eff}}}{(a^1)^{1+\omega_d^{eff}}}+\frac{4\pi}{3}(R_A^1)^3\Delta \rho_d\right]
\end{eqnarray} 
When $\Delta t=t_1-t_0 \rightarrow \mathrm{d}t$, eq.(\ref{eq28}) could be generalized to
\begin{equation}\label{eq29}
\mathrm{d}E_A=\frac{4\pi}{3}(R_A^1)^3\rho^0-\frac{4\pi}{3}(R_A^0)^3\left[\rho^0_m\frac{(a^0)^{1+\omega_m^{eff}}}{(a^1)^{1+\omega_m^{eff}}}+\rho^0_d\frac{(a^0)^{1+\omega_d^{eff}}}{(a^1)^{1+\omega_d^{eff}}}\right]+\frac{4\pi}{3}(R_A^1)^3\dot{\rho}\mathrm{d}t
\end{equation}
\indent On the other hand, it has been found that black holes emit Hawking radiation 
with temperature proportional to their surface gravity at the event horizon 
and they have entropy which is one quarter of the area of the event horizon 
in the semiclassical quantum description of black hole physics \cite{f38}. 
If we define the temperature and entropy on the apparent horizon as
\begin{equation}\label{eq30}
T_A=\frac{1}{2\pi R_A},\ \
S_A=\frac{A}{4G}=\frac{\pi R_A^2}{G}
\end{equation}
At $t=t_1$, $R_A^0 \rightarrow R_A^1$ and $S_A \rightarrow S_A^1$, 
taking differential form of the above equation and using eq.(\ref{eq12}), 
we can obtain
\begin{equation}\label{eq31}
T_A^1\mathrm{d}S_A=\frac{\dot{R_A}}{G}\mathrm{d}t=-\frac{4\pi}{3}(R_A^1)^3\dot{\rho}\mathrm{d}t
\end{equation}
Here we have used $T_A\approx T_A^1$ (quasi-equilibrium process). 
Substituting this relation into eq.(\ref{eq29}), 
we get the modified first law of thermodynamics on the apparent horizon
\begin{equation}\label{eq32}
-\mathrm{d}E_A=-\left\{\frac{4\pi}{3}(R_A^1)^3\rho^0-\frac{4\pi}{3}(R_A^0)^3\left[\rho^0_m\frac{(a^0)^{1+\omega_m^{eff}}}{(a^1)^{1+\omega_m^{eff}}}+\rho^0_d\frac{(a^0)^{1+\omega_d^{eff}}}{(a^1)^{1+\omega_d^{eff}}}\right]\right\}+T_A^1\mathrm{d}S_A
\end{equation}
so we can write 
\begin{equation}\label{eq33}
\mathrm{d}W=-\left\{\frac{4\pi}{3}(R_A^1)^3\rho^0-\frac{4\pi}{3}(R_A^0)^3\left[\rho^0_m\frac{(a^0)^{1+\omega_m^{eff}}}{(a^1)^{1+\omega_m^{eff}}}+\rho^0_d\frac{(a^0)^{1+\omega_d^{eff}}}{(a^1)^{1+\omega_d^{eff}}}\right]\right\}
\end{equation}
This expression is a little more complicated but if we make a constraint to the parameters 
the classic form of the first law of thermodynamics would be obtained through this formula. 
The adjustable parameters might make $\omega_m^{eff}$ and $\omega_d^{eff}$ approaching $-1$, 
under this condition
\begin{equation}\label{eq34}
\mathrm{d}W=-\frac{4\pi}{3}\rho^0\left[(R_A^1)^3-(R_A^0)^3\right]=-4\pi(R_A^0)^2\rho_d^0(1+r)\mathrm{d}R_A
\end{equation}
And eq.(\ref{eq32}) can be modified as
\begin{equation}\label{eq35}
-\mathrm{d}E_A=\mathrm{d}W+T_A^1\mathrm{d}S_A=-4\pi(R_A^0)^2\rho_d^0(1+r)\mathrm{d}R_A-\frac{4\pi}{3}(R_A^1)^3\dot{\rho}\mathrm{d}t
\end{equation}
Here the work density term $\mathrm{d}W$ \cite{f39} is regarded as the work done 
by the change of the apparent horizon, which is used to replace the negative pressure 
if compared with the standard first law of thermodynamics. 
It is not difficult to understand that for the other spatial curvature 
the conclusion is also hold except the difference of $\Omega_k$.

\section{GENERALIZED SECOND LAW OF THERMODYNAMICS ON THE APPARENT HORIZON}

\indent Now we proceed to discuss the generalized second law of 
thermodynamics in a region enclosed by the apparent horizon. 
Using eq.(\ref{eq33}) we turn to find out
\begin{equation}\label{eq36}
T_A^1\dot{S_A}=-\frac{4\pi}{3}(R_A^1)^3\dot{\rho}
\end{equation}
\indent To check the generalized second law of thermodynamics, 
we have to examine the evolution of the total entropy. 
In order to obtain the variation of the entropy of the fluid inside 
the apparent horizon we use the Gibb's equation \cite{f40}
\begin{equation}\label{eq37}
T\Delta S=\Delta E+p_d\Delta V
\end{equation}
Here the total entropy of the energy and matter content inside the horizon is $S=S_m+S_d$. 
Since the interaction of both dark components is mutual, we assumed that the temperatures of 
them are equal and it should be in equilibrium with the temperature associated 
with the apparent horizon $T_A=T_m=T_d$.\\
\indent In the model discussed above, the comoving distance has a faster expansion than 
the apparent horizon radius and there should be some energy flow through the apparent horizon for the static observer in space. 
Therefore from the Gibb's equation (\ref{eq37}) we can obtain
\begin{equation}\label{eq38}
T_A^1\Delta S=\rho^1\Omega_k\left[(R_A^1)^3-(a^1\xi)^3\right]+p_d\Omega_k\left[(R_A^1)^3-(R_A^0)^3\right]
\end{equation}
Where $p_d^0 \approx p_d^1$, $T_A \approx T_A^1$, 
if the quasi-equilibrium process is taken into account. For spatially flat case
\begin{eqnarray}\label{eq39}
T_A^1(\Delta S_m+\Delta S_d)&=&\left[\frac{4\pi}{3}(R_A^1)^3\rho^0_m-\frac{4\pi}{3}(a^0\xi)^3\rho^0_m\frac{(a^0)^{1+\omega_m^{eff}}}{(a^1)^{1+\omega_m^{eff}}}+\frac{4\pi}{3}(R_A^1)^3\Delta \rho_m\right] \notag\\
&+&\left[\frac{4\pi}{3}(R_A^1)^3\rho^0_d-\frac{4\pi}{3}(a^0\xi)^3\rho^0_d\frac{(a^0)^{1+\omega_d^{eff}}}{(a^1)^{1+\omega_d^{eff}}}+\frac{4\pi}{3}(R_A^1)^3\Delta \rho_d\right]\notag\\
&+&\omega_d\rho_d^0\frac{4\pi}{3}\left[(R_A^1)^3-(R_A^0)^3\right]
\end{eqnarray} 
Making use of the condition $\Delta t=t_1-t_0 \rightarrow \mathrm{d}t$, 
$\omega_d^{eff} \rightarrow -1$ and $\omega_m^{eff} \rightarrow -1$, 
eq.(\ref{eq39}) can be rewritten as
\begin{eqnarray}\label{eq40}
T_A^1(\mathrm{d}S_m+\mathrm{d}S_d)&=&\left\{\frac{4\pi}{3}\rho^0_m\left[(R_A^1)^3-(R_A^0)^3\right]+\frac{4\pi}{3}(R_A^1)^3 \mathrm{d}\rho_m\right\}\notag\\
&+&\left\{\frac{4\pi}{3}\rho^0_d\left[(R_A^1)^3-(R_A^0)^3\right]+\frac{4\pi}{3}(R_A^1)^3 \mathrm{d}\rho_d\right\}\notag\\
&+&\omega_d\rho_d^0\frac{4\pi}{3}\left[(R_A^1)^3-(R_A^0)^3\right]
\end{eqnarray}
or
\begin{equation}\label{eq41}
T_A^1(\mathrm{d}S_m+\mathrm{d}S_d)=\left\{\frac{4\pi}{3}\rho^0\left[(R_A^1)^3-(R_A^0)^3\right]+\frac{4\pi}{3}(R_A^1)^3 \dot{\rho}\mathrm{d}t\right\}+\omega_d\rho_d^0\frac{4\pi}{3}\left[(R_A^1)^3-(R_A^0)^3\right]
\end{equation}
which can be recast as
\begin{equation}\label{eq42}
T_A^1(\dot{S_m}+\dot{S_d})=\rho_d^0(1+r+\omega_d)4\pi (R_A^0)^2\dot R_A+\frac{4\pi}{3}(R_A^1)^3\dot{\rho}
\end{equation}
Hence combining eqs.(\ref{eq36}) and (\ref{eq42}) the resulting change of total entropy is given by
\begin{equation}\label{eq43}
T_A^1(\dot{S_m}+\dot{S_d}+\dot{S_A})=\rho_d^0(1+r+\omega_d)4\pi (R_A^0)^2\dot R_A
\end{equation}
Substituting eq.(\ref{eq12}) into (\ref{eq43})
\begin{equation}\label{eq43}
T_A^1(\dot{S_m}+\dot{S_d}+\dot{S_A})=16\pi^2GH(R_A^0)^5(\rho_d^0)^2(1+r+\omega_d)^2
\end{equation}
It is the total entropy of the universe and the apparent horizon entropy. 
Therefore, the right hand side of the above equation is always positive in the evolution of the universe. 
This indicates that for a universe with special curvature filled with interacting dark components, 
the generalized second law of thermodynamics is fulfilled in a region enclosed by the apparent horizon.

\section{conclusion}

\indent In this paper, we have studied the thermodynamical properties of the universe with dark energy. 
It is demonstrated that in a universe with spacial curvature the natural choice for 
IR cutoff could be the apparent horizon radius. The apparent horizon is important for the study of cosmology, 
since on the apparent horizon there is the well known correspondence between 
the first law of thermodynamics and the Einstein equation. On the other hand it has been 
found that the apparent horizon is a good boundary for keeping thermodynamical laws. \\
\indent We showed that any interaction of pressureless dark matter with HDE, 
whose infrared cutoff is set by the apparent horizon radius, implying a constant 
effective equation of state of dark component in a universe with any spatial curvature. 
In addition, motivated by the idea in \cite{f33}, we also performed that for the static observer in space, 
the comoving distance has a faster expansion than the apparent horizon radius. 
Then we discussed the validity of the first and the generalized second law of thermodynamics. 
In the condition of $\omega_d^{eff} \rightarrow -1$ and $\omega_m^{eff} \rightarrow -1$, 
both of them could return to the classic form and they can be always satisfied 
for a universe filled with mutual interacting dark components in a region enclosed by the apparent horizon.\\
\indent The results tell us that in studying the thermodynamics of the universe with dark energy, 
it is more appropriate to consider the difference between apparent horizon radius and the comoving distance. 
Our study further supports that in a universe with spatial curvature, 
the apparent horizon is a physical boundary from the thermodynamical point of view.

\section{ackonwledgement}

The authors are grateful to Yu-Xiao Liu, Yong-Qiang Wang, Shao-Wen Wei and Ke Yang for helpful discussions.

\newpage
\begin{figure}
\begin{center}
\includegraphics[height=6.5cm]{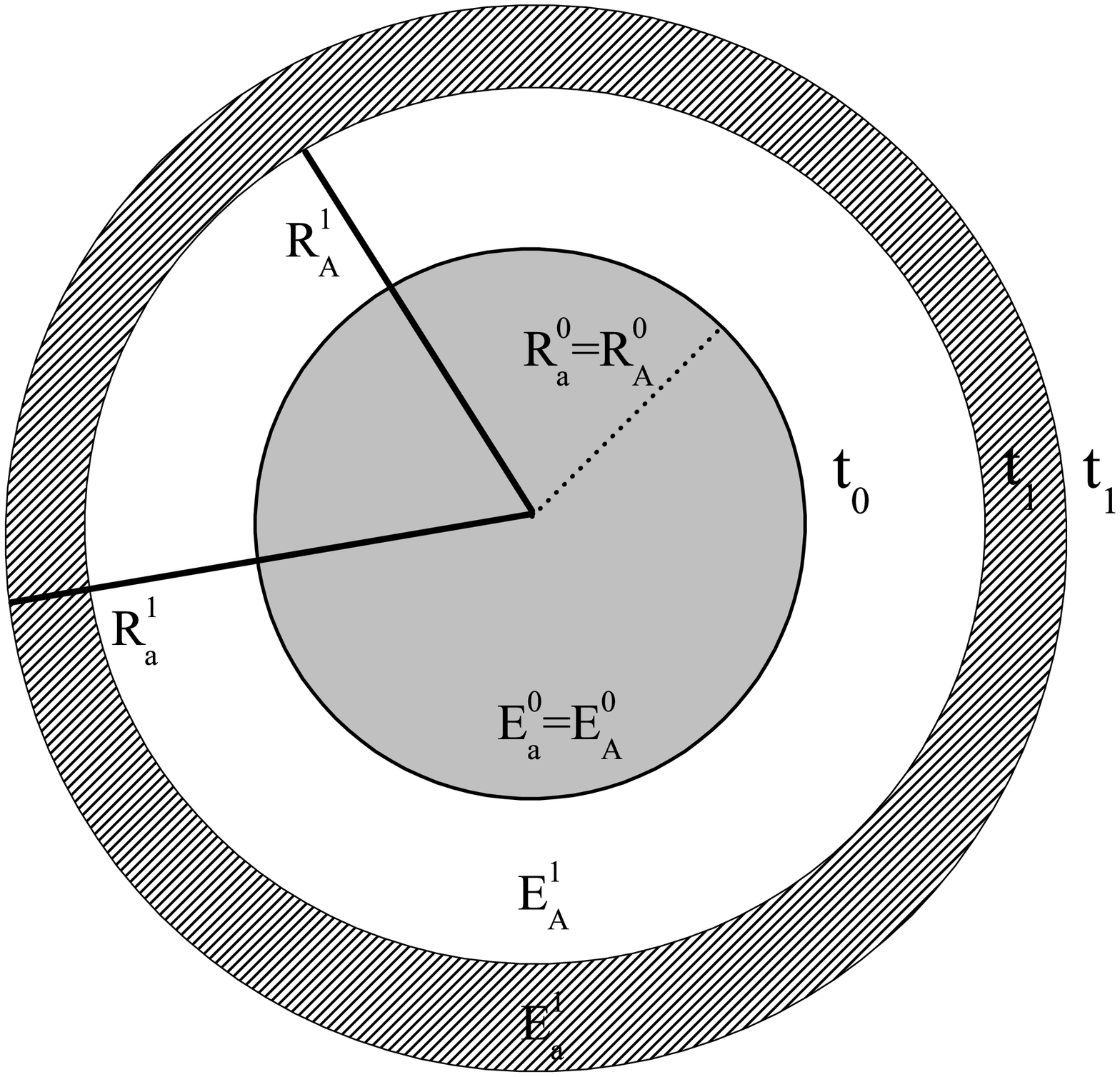}    
\caption{The evolution of the apparent horizon and commoving distance with scale factor $a$ is given by eq.(\ref{eq24}). at a certain beginning time, $t=t_0$, $a^0 \xi=R_A^0$. after a time $\Delta t$, $t_0 \rightarrow t_1$, $R_a^1>R_A^1$.} \label{fig:bifurcation}
\end{center}
\end{figure}

\end{document}